\documentstyle[12pt]{article}
\title{ An Algorithm for the Constrained Longest Common Subsequence and Substring Problem }
\author {Rao Li \\ 
          Dept. of Computer Science, Engineering, and Mathematics \\
         University of South Carolina Aiken \\
	   Aiken, SC 29801 \\
           USA \\
         {\it Email: raol@usca.edu } \and
Jyotishmoy Deka \\
Dept. of Electrical Engineering \\
Tezpur University \\
Tezpur, Assam 784028 \\
India \\
{\it Email: jyotishmoydeka62@gmail.com}  \and 
 Kaushik Deka \\
Dept. of Computer Science and Engineering \\
National Institute of Technology Silchar \\
Cachar, Assam 788010 \\
India \\ 
{\it Email: jagatdeka20@gmail.com} \and
Dorothy Li \\
12000 Market Street, Unit 63 \\
Reston, VA 20190 \\
USA \\ 
{\it Email: dorothy.li1994@gmail.com}
         }
\date{Aug. 1, 2023}
\begin{document}
\maketitle

\begin{abstract}
Let $\Sigma$ be an alphabet. 
For two strings $X$, $Y$, and a constrained string $P$ over the alphabet $\Sigma$, 
the constrained longest common subsequence and substring problem for two strings $X$ and $Y$ with respect to $P$ 
is to find a longest string $Z$ which is a subsequence of $X$, a substring of $Y$, and has $P$ as a subsequence. 
In this paper, we propose an algorithm for the constrained longest common subsequence and substring problem for two strings 
with a constrained string.
 \end{abstract}
\vspace{2mm}

\hspace{2mm} Keywords: longest common subsequence,  longest \newline  
    \hspace*{29mm}  common  substring, longest common \newline   
  \hspace*{29mm}   subsequence and substring, constrained  \newline  
  \hspace*{29mm}   longest common subsequence \newline

\noindent {\bf 1.  Introduction} \\

Let $\Sigma$ be an alphabet and $S$ a string over $\Sigma$. A subsequence of a string $S$ over an alphabet $\Sigma$ is obatined by deleting zero or more letters of $S$.
A substring of a string $S$ is a subsequence of $S$ consists of consecutive letters in $S$.
The longest common subsequence problem (LCSSeq) for two strings is to find a longest string which is a subsequence of both strings. 
The longest common substring (LCSStr) problem for two strings is to find a longest string which is a substring of both strings. 
Both the longest common subsequence problem and the longest common substring problem
have been well-studied in last several decades. 
More details on the studies for the first problem
can be found in \cite{Apo}, \cite {Apo1}, \cite{Bergroth}, \cite{Cormen}, \cite{H}, \cite{Hberg}, \cite{Hunt},
and \cite{Rick} and the second problem can be found in \cite{G} and \cite{W}.  \newline

Tsai \cite{Tsai} extended the  longest common subsequence problem for two strings to the 
constrained longest common subsequence (CLCSSeq) probelm
for two strings. For two strings $X$, $Y$, and a constrained string $P$, 
the constrained longest common subsequence problem for two strings $X$ and $Y$ with respect to $P$ 
is to find a string $Z$ such that $Z$ is a longest 
common subsequence for $X$ and $Y$ and $P$ is a subsequence of $Z$. 
Tsai \cite{Tsai} designed an $O(|X|^2 |Y|^2 |P|)$ time algorithm for the CLCSSeq problem for two strings,
where $|X|$, $|Y|$, and $|P|$ denote the lengths of the strings $X$, $Y$, and $P$, respectively. 
Chin et al. \cite{Chin} improved Tsai's algorithm and designed an $O(|X| |Y| |P|)$ time algorithm 
for the CLCSSeq problem for two strings $X$ and $Y$ and a constrained string $P$. \\

Motivated by LCSSeq and LCSStr problems, Li et. al \cite{LDD} introduced the longest common subsequence and substring (LCSSeqSStr) problem for two strings. 
For two strings $X$, $Y$, the longest common subsequence and substring problem for $X$ and $Y$ is to find a longest string which is a subsequence of $X$ and a substring of $Y$. 
They also designed an $O(|X||Y|)$ time algorithm for LCSSeqSStr problem for two strings $X$ and $Y$ in \cite{LDD}.  \\

Motivated by Tsai's extension of LCSSeq to CLCSSeq for two strings, we introduce the 
 constrained longest common subsequence and substring problem for two strings with respect to a constrained string. 
For two strings $X$, $Y$, and a constrained string $P$, 
the constrained longest common subsequence and substring (CLCSSeqSStr) problem for two strings $X$ and $Y$ with respect to $P$ 
is to find a string $Z$ such that $Z$ is a longest common subsequence of $X$, a substring of $Y$, and has $P$ as a subsequence. 
Clearly, the CLCSSeq problem is a special CLCSSeqSStr problem with an empty constrained string.  
In this paper, we, using some ideas and techniques developed in \cite{Chin}, design an $O(|X| |Y| |P|)$ time algorithm for 
CLCSSeqSStr problem for two strings and a constrained string. \newline

\noindent {\bf 2.  The Recursions in the Algorithm} \\

In order to present our algorithm, we need to establish some recursions to be used in our algorithm. Before establishing the recursions, we need some notations as follows.
For a given string $S = s_1 s_2 ... s_l$ over an alphabet $\Sigma$, the size of $S$, denoted $|S|$, is defined as the number of letters in $S$.
The $i$ prefix of $S$ is defined as $S_i = s_1 s_2 ... s_i$, where $1 \leq i \leq l$. Conventionally, $S_0$ is defined as an empty string.
The $l$ suffixes of $S$ are the strings of $s_1 s_2 ... s_l$, $s_2 s_3 ... s_l$, ..., $s_{l - 1}s_l$, and $s_l$. Let $X = x_1 x_2 ... x_m$ and
 $Y = y_1 y_2 ... y_n$ be two strings and  $P = p_1 p_2 ... p_r$ a constrained string. 
We define $Z[i, j, k]$ as a string satisfying the following conditions, where $1 \leq i \leq m$, $1 \leq j \leq n$, and $1 \leq k \leq r$, \\  \\
 \hspace*{7mm} ($1$) it is a subsequence of $X_i$, \newline
 \hspace*{7mm} ($2$) it is a suffix of $Y_j$, \newline
\hspace*{7mm} ($3$) it has $P_k$ as a subsequence, \newline
 \hspace*{7mm} ($4$) under ($1$), ($2$) and (3), its length is as large as possible. \\ 
 
\noindent{\bf Claim $1$.} Let $U^k = u_1^k u_2^k ... u_{h_k}^k$ be a longest string which is a subsequence of $X$, a substring of $Y$, and has $P_k$ as a subsequence. 
Then $h_k = \max \{\, |Z[i, j, k]| : 1 \leq i \leq m,  1 \leq j \leq n, 1 \leq k \leq r \,\}$. \\

\noindent {\bf Proof of Claim $1$.} For each $i$ with $1 \leq i \leq m$, each $j$ with $1 \leq j \leq n$, and each $k$ with $1 \leq k \leq r$, we, from the definition of $Z[i, j, k]$, have that $Z[i, j, k]$ is a subsequence of $X$, a substring of $Y$, and has $P_k$ as a subsequence.
By the definition of $U^k$, we have that $|Z[i, j, k]| \leq |U^k| = h_k$. Thus $\max \{\, |Z[i, j, k]| : 1 \leq i \leq m,  1 \leq j \leq n, 1 \leq k \leq r  \,\} \leq h_k$. \\

Since $U^k = u_1^k u_2^k ... u_{h_k}^k$ is a longest string which is a subsequence of $X$, a substring of $Y$, and has $P_k$ as a subsequence, there is an index $s$ and an index $t$ such that $u_{h_k}^k = x_s$ and $u_{h_k}^k = y_t$ such that
$U^k = u_1^k u_2^k ... u_{h_k}^k$ is a subsequence of $X_s$, a suffix of $Y_t$, and  has $P_k$ as a subsequence. From the definition of $Z[i, j, k]$, we have that $h_k \leq |Z[s, t, k]| \leq \max \{\, |Z[i, j, k]| : 1 \leq i \leq m,  1 \leq j \leq n, 1 \leq k \leq r \,\}$.  \\

Hence $h_k = \max \{\, |Z[i, j, k]| : 1 \leq i \leq m,  1 \leq j \leq n, 1 \leq k \leq r \,\}$ and the proof of Claim $1$ is complete. \\

\noindent{\bf Claim $2$.} Suppose that $X_i = x_1 x_2 ... x_i$, $Y_j = y_1 y_2 ... y_j$, and $P = p_1 p_2 ... p_k$, where $1 \leq i \leq m$ and $1 \leq j \leq n$, $1 \leq k \leq r$. 
If $Z[i, j, k] = z_1 z_2 ... z_a$ is a string satisfying conditions ($1$), ($2$), ($3$), and ($4$) above. Then we have only the following possible cases and the statement in each case is true. \\

\noindent {\bf Case $1$}. $x_i = y_j = p_k$. We have $|Z[i, j, k]| = |Z[i - 1, j - 1, k - 1]| + 1$ in this case.  \\

\noindent {\bf Case $2$}.  $x_i = y_j \neq p_k$. We have $|Z[i, j, k]| = |Z[i - 1, j - 1, k]| + 1$ in this case.  \\

\noindent {\bf Case $3$}. $x_i \neq y_j $, $x_i \neq p_k$, and $y_j = p_k$. We have $|Z[i, j, k]| = |Z[i - 1, j, k||$ in this case.  \\

\noindent {\bf Case $4$}. $x_i \neq y_j $, $x_i \neq p_k$, and $y_j \neq p_k$. We have $|Z[i, j, k]| = |Z[i - 1, j, k||$ in this case.  \\

\noindent {\bf Case $5$}. $x_i \neq y_j $, $x_i = p_k$, and $y_j \neq p_k$. This case does not happen.  \\

\noindent{\bf Proof of Claim $2$.} The five cases can be figured out in the following way. Firstly, we have two cases of $x_i = y_j$ or  $x_i \neq y_j$. When $x_i = y_j$, we just can have two possible cases of 
 $x_i = y_j = p_k$ or $x_i = y_j \neq p_k$. When  $x_i \neq y_j$, we just can have three possible cases of $x_i \neq p_k$ and $y_j = p_k$, $x_i \neq p_k$ and $y_j \neq p_k$, or $x_i = p_k$ and $y_j \neq p_k$. 
Next we will prove the statements in the five cases. \\

\noindent {\bf Case $1$.} Since $Z[i, j, k] = z_1 z_2 ... z_a$ is a suffix of $Y_j$, we have that $z_a = y_j = x_i = p_k$. Let $W = w_1w_2 ... w_b = Z[i - 1, j - 1, k - 1]$ be a string satisfying the following conditions, \\  \\
 \hspace*{7mm} - it is a subsequence of $X_{i - 1}$. \newline
 \hspace*{7mm} - it is a suffix of $Y_{j - 1}$, \newline
\hspace*{7mm} - it has $P_{k - 1}$ as a subsequence, \newline
 \hspace*{7mm} - under ($1$), ($2$) and (3), its length is as large as possible. \\  \\
Note that $z_1 z_2 ... z_{a - 1}$ is a string which is a subsequence of $X_{i - 1}$, a suffix of $Y_{j - 1}$, and has $P_{k - 1}$ as a subsequence. By the definition of $W = w_1w_2 ... w_b$, we have that $a - 1 \leq b$. Namely, $a \leq b + 1$. \\

\noindent Note that $w_1w_2 ... w_bz_a$ is a string satisfying following conditions, \\  \\
 \hspace*{32mm} - it is a subsequence of $X_i$, \newline
 \hspace*{32mm} - it is a suffix of $Y_j$, \newline
\hspace*{32mm} - it has $P_k$ as a subsequence. \\  \\
By the definition of $Z[i, j, k] = z_1 z_2 ... z_a$, we have that $b + 1 \leq a$. Thus $a = b + 1$ and $|Z[i, j, k]| = |Z[i - 1, j - 1, k - 1]| + 1$. \\ 

\noindent {\bf Case $2$.}   Since $Z[i, j, k] = z_1 z_2 ... z_a$ is a suffix of $Y_j$, we have that $z_a = y_j = x_i \neq p_k$. Let $U = u_1u_2 ... u_c = Z[i - 1, j - 1, k]$ be a string satisfying the following conditions, \\  \\
 \hspace*{7mm} - it is a subsequence of $X_{i - 1}$, \newline
 \hspace*{7mm} - it is a suffix of $Y_{j - 1}$, \newline
\hspace*{7mm} - it has $P_{k }$ as a subsequenc, \newline
 \hspace*{7mm} - under ($1$), ($2$) and (3), its length is as large as possible. \\  \\
Note that $z_1 z_2 ... z_{a - 1}$ is a string which is a subsequence of $X_{i - 1}$, a suffix of $Y_{j - 1}$, and has $P_{k }$ as a subsequence. By the definition of $U = u_1u_2 ... u_c = Z[i - 1, j - 1, k]$, we have that $a - 1 \leq c$. Namely, $a \leq c + 1$. \\

\noindent Note that $u_1u_2 ... u_c$ is a string satisfying the following conditions, \\ \\
 \hspace*{32mm} - it is a subsequence of $X_{i - 1}$, \newline
 \hspace*{32mm} - it is a suffix of $Y_{j  - 1}$,\newline
\hspace*{32mm} - it has $P_k$ as a subsequence. \\ \\
Thus $u_1u_2 ... u_cy_j$ is a string which is a subsequence of $X_i$,  a suffix of $Y_j$, and has $P_k$ as a subsequence. By the definition of $Z[i, j, k] = z_1 z_2 ... z_a$, we have that $c + 1 \leq a$. Thus $a = c + 1$ and $|Z[i, j, k]| = |Z[i - 1, j - 1, k ]| + 1$. \\ 

\noindent {\bf Case $3$.}   Since $Z[i, j, k] = z_1 z_2 ... z_a$ is a suffix of $Y_j$, we have that $z_a = y_j = p_k \neq x_i$. Let $V = v_1v_2 ... v_d = Z[i - 1, j, k]$ be a string satisfying the following conditions, \\  \\
 \hspace*{7mm} - it is a subsequence of $X_{i - 1}$, \newline
 \hspace*{7mm} - it is a suffix of $Y_{j }$, \newline
\hspace*{7mm} - it has $P_{k }$ as a subsequence, \newline
 \hspace*{7mm} - under ($1$), ($2$) and (3), its length is as large as possible. \\  \\
Note that $z_1 z_2 ... z_{a}$ is a string which is a subsequence of $X_{i - 1}$, a suffix of $Y_{j }$, and has $P_{k }$ as a subsequence. By the definition of $V = v_1v_2 ... v_d = Z[i - 1, j, k]$, we have that $a  \leq d$.  \\

\noindent Note that $v_1v_2 ... v_d$ is a string satisfying conditions, \\ \\
 \hspace*{32mm} - it is a subsequence of $X_{i - 1}$, \newline
 \hspace*{32mm} - it is a suffix of $Y_{j }$, \newline
\hspace*{32mm} - it has $P_k$ as a subsequence. \\ \\
Thus $v_1v_2 ... v_d$ is a string which is a subsequence of $X_i$,  a suffix of $Y_j$, and has $P_k$ as a subsequence. By the definition of $Z[i, j, k] = z_1 z_2 ... z_a$, we have that $d \leq a$. Thus $a = d$ and $|Z[i, j, k]| = |Z[i - 1, j, k ]|$. \\ 

\noindent {\bf Case $4$.}   Since $Z[i, j, k] = z_1 z_2 ... z_a$ is a suffix of $Y_j$, we have that $z_a = y_j \neq p_k$, $z_a = y_j \neq x_i$, and $x_i \neq p_k$. 
Let $Q = q_1q_2 ... q_e = Z[i - 1, j, k]$ be a string satisfying the following conditions, \\  \\
 \hspace*{7mm} - it is a subsequence of $X_{i - 1}$, \newline
 \hspace*{7mm} - it is a suffix of $Y_{j }$, \newline
\hspace*{7mm} - it has $P_{k }$ as a subsequence, \newline
 \hspace*{7mm} - under ($1$), ($2$) and (3), its length is as large as possible. \\  \\
Note that $z_1 z_2 ... z_{a}$ is a string which is a subsequence of $X_{i - 1}$, a suffix of $Y_{j }$, and has $P_{k }$ as a subsequence. By the definition of $Q = q_1q_2 ... q_e = Z[i - 1, j, k]$, we have that $a  \leq e$.  \\

\noindent Note that $q_1q_2 ... q_e$ is a string satisfying the following conditions, \\ \\
 \hspace*{32mm} - it is a subsequence of $X_{i - 1}$, \newline
 \hspace*{32mm} - it is a suffix of $Y_{j }$, \newline
\hspace*{32mm} - it has $P_k$ as a subsequence. \\ \\
Thus $q_1q_2 ... q_e$ is a string which is a subsequence of $X_i$,  a suffix of $Y_j$, and has $P_k$ as a subsequence. By the definition of $Z[i, j, k] = z_1 z_2 ... z_a$, we have that $e \leq a$. Thus $a = e$ and $|Z[i, j, k]| = |Z[i - 1, j, k ]|$. \\ 

\noindent {\bf Case $5$}.  Since $Z[i, j, k] = z_1 z_2 ... z_a$ is a suffix of $Y_j$, we have that $z_a = y_j \neq x_i  = p_k$. Since $z_1 z_2 ... z_a$ is a subsequence of $X_i$ and $x_i \neq z_a$, we have that $z_a$ appears before $x_i$ on $X_i$. 
Since $x_i = p_k$ on $X_i$, $p_1p_2 ... p_k$ cannot be a subsequence of $z_1z_2 ... z_a$, a conradiction. 
Note that since this case does not happen, we will not deal with this case in our algorithm. \\ 

Therefore the proof of Claim $2$ is complete. \\

The following Claim $3$ which will used in our algorithm demonstrates the implications of the condition that there is not a string which is a subsequence of $X_i = x_1 x_2 ... x_i$, a suffix of 
$Y_j = y_1 y_2 ... y_j$, and has $P_k = p_1 p_2 ... p_k$ as a subsequence.  \newline

\noindent{\bf Claim $3$.} Suppose there is not a string which is a subsequence  of $X_i = x_1 x_2 ... x_i$, a suffix of 
$Y_j = y_1 y_2 ... y_j$, and has $P_k = p_1 p_2 ... p_k$ as a subsequence.  \\

\noindent  [$1$]. If $x_i = y_j = p_k$, then there is not a string which is a subsequence  of $X_{i - 1} = x_1 x_2 ... x_{i - 1}$, a suffix of 
$Y_{j - 1} = y_1 y_2 ... y_{j - 1} $, and has $P_{k - 1} = p_1 p_2 ... p_{k - 1}$ as a subsequence.  \\

\noindent [$2$].  If $x_i = y_j \neq p_k$, then there is not a string which is a subsequence  of $X_{i - 1} = x_1 x_2 ... x_{i - 1}$, a suffix of 
$Y_{j - 1} = y_1 y_2 ... y_{j - 1} $, and has $P_k = p_1 p_2 ... p_{k}$ as a subsequence.  \\

\noindent [$3$]. If $x_i \neq y_j $, $x_i \neq p_k$, and $y_j = p_k$, then there is not a string which is a subsequence of $X_{i - 1} = x_1 x_2 ... x_{i - 1}$, a suffix of 
$Y_j = y_1 y_2 ... y_{j } $, and has $P_k = p_1 p_2 ... p_{k}$ as a subsequence.  \\

\noindent [$4$]. If $x_i \neq y_j $, $x_i \neq p_k$, and $y_j \neq p_k$, then there is not a string which is a subsequence  for $X_{i - 1} = x_1 x_2 ... x_{i - 1}$, a suffix of 
$Y_j = y_1 y_2 ... y_{j } $, and has $P_k = p_1 p_2 ... p_{k}$ as a subsequence.  \\


\noindent{\bf Proof of Claim $3$.} We next will prove the statements in the four cases. \\

\noindent [$1$]. Now we have that $x_i = y_j = p_k$. Suppose, to the contrary, that there is a string $W_1$ which is a subsequence  of $X_{i - 1} = x_1 x_2 ... x_{i - 1}$, a suffix of 
$Y_{j - 1} = y_1 y_2 ... y_{j - 1} $, and has $P = p_1 p_2 ... p_{k - 1}$ as a subsequence. Then $W_1x_i$ is a string which is a subsequence of $X_i = x_1 x_2 ... x_i$, a suffix of 
$Y_j = y_1 y_2 ... y_j$, and has $P_k = p_1 p_2 ... p_k$ as a subsequence, a contradiction. \\

\noindent [$2$].  Now we have that $x_i = y_j \neq p_k$. Suppose, to the contrary, that there is a string $W_2$ which is a subsequence of $X_{i - 1} = x_1 x_2 ... x_{i - 1}$, a suffix of 
$Y = y_1 y_2 ... y_{j - 1} $, and has $P_k = p_1 p_2 ... p_{k}$ as a subsequence. Then $W_2x_i$ is a string which is a subsequence  of $X_i = x_1 x_2 ... x_i$, a suffix of 
$Y_j = y_1 y_2 ... y_j$, and has $P_k = p_1 p_2 ... p_k$ as a subsequence, a contradiction. \\

\noindent [$3$]. Now we have that $x_i \neq y_j $, $x_i \neq p_k$, and $y_j = p_k$. Suppose, to the contrary, that there is a string $W_3$ which is a subsequence  for $X_{i - 1} = x_1 x_2 ... x_{i - 1}$, a suffix of 
$Y_j = y_1 y_2 ... y_{j } $, and has $P_k = p_1 p_2 ... p_{k}$ as a subsequence. Then $W_3$ is a string which is a subsequence  of $X_i = x_1 x_2 ... x_i$, a suffix of 
$Y_j = y_1 y_2 ... y_j$, and has $P_k = p_1 p_2 ... p_k$ as a subsequence, a contradiction. \\

\noindent [$4$].  Now we have that $x_i \neq y_j $, $x_i \neq p_k$, and $y_j \neq p_k$. Suppose, to the contrary, that there is a string $W_4$ which is a subsequence of $X_{i - 1} = x_1 x_2 ... x_{i - 1}$, a suffix of 
$Y_j = y_1 y_2 ... y_{j } $, and has $P_k = p_1 p_2 ... p_{k}$ as a subsequence. Then $W_4$ is a string which is a subsequence  of $X_i = x_1 x_2 ... x_i$, a suffix of 
$Y = y_1 y_2 ... y_j$, and has $P_k = p_1 p_2 ... p_k$ as a subsequence, a contradiction. \\

Therefore the proof of Claim $3$ is complete. \\

\noindent {\bf 3.  The Algorithm} \\

Now we can present our algorithm. We assume that $X = x_1x_2 ... x_m$, $Y = y_1y_2 ... y_n$, and $P = p_1p_2 ... p_r$. Let $M$ be a three dimensional array of size $(m + 1)(n + 1)(r + 1)$. It can be thought as a collection
of $(r + 1)$ two dimensional arrays of size $(m + 1)(n + 1)$.
The cells $M[i][j][k]$,  where $0 \leq i \leq m$, $0 \leq j \leq n$, and $0 \leq k \leq r$, store the lengths of longest strings such that each of them is a subsequence of $X_i$, a suffix of $Y_j$, and has $P_k$ as a subsequence. \\
If either $i < r$ or $j < r$, there is not a string which is a subsequence of $X_i$, a suffix of $Y_j$, and has $P_k$ as a subsequence. This situation is represented by setting $M[i][j][k] = -\infty$, where $\infty$ can be any number which 
is greater than the larger one between $m$ and $n$. Now we can fill in some boundary cells in array $M$. \\

\noindent If $i = 0$ and $k = 0$ or $j = 0$ and $k = 0$, the length of a string which is a subsequence of $X_i$, a suffix of $Y_j$, and has $P_k$ as a subsequence is zero. Thus $M[0][j][0] = 0$, where $0 \leq j \leq n$ and 
$M[i][0][0] = 0$, where $0 \leq i \leq m$. \\

\noindent If $k = 0$ or $P$ is an empty string. The CLCSSeqSStr problem for two strings $X$ and $Y$ and a constrained string $P$ becomes the LCSSeqSStr problem for two strings $X$ and $Y$. The cells of
$M[i][j][0]$, where $1 \leq i \leq m$ and $1 \leq j \leq n$, can be filled in by the following rules. If $x_i = y_j$, then $M[i][j] = M[i - 1][j - 1] + 1$. If $x_i \neq y_j$, then $M[i][j] = M[i - 1][j]$. The detailed proofs for the truth
of the rules can be found in \cite{LDD}. \\

\noindent If $i = 0$ and $k \geq 1$, there is not a string which is a subsequence of $X_i$, a suffix of $Y_j$, and has $P_k$ as a subsequence. Thus $M[0][j][k] = - \infty$, where $0 \leq j \leq n$ and $1 \leq k \leq r$. \\

\noindent If $j = 0$ and $k \geq 1$, there is not a string which is a subsequence of $X_i$, a suffix of $Y_j$, and has $P$ as a subsequence. Thus $M[i][0][k] = - \infty$, where $0 \leq i \leq m$ and $1 \leq k \leq r$. \\

Next we will fill in the remaining cells $M[i][j][k]$, where $i \geq 1$, $j \geq 1$,  and $k \geq 1$. \\

\noindent If $i \geq 1$, $j \geq 1$,  $k \geq 1$, and $x_i = y_j = p_k$, then $M[i][j][k] =  M[i - 1][j - 1][k - 1] + 1$. \\

\noindent If $i \geq 1$, $j \geq 1$,  $k \geq 1$, and $x_i = y_j \neq p_k$, then $M[i][j][k] =  M[i - 1][j - 1][k ] + 1$. \\

\noindent If $i \geq 1$, $j \geq 1$,  $k \geq 1$, and $x_i \neq y_j$, $x_i \neq p_k$, and $y_j = p_k$, then $M[i][j][k] =  M[i - 1][j][k ]$. \\

\noindent If $i \geq 1$, $j \geq 1$,  $k \geq 1$, and $x_i \neq y_j$, $x_i \neq p_k$, and $y_j \neq p_k$, then $M[i][j][k] =  M[i - 1][j][k ]$. \\

Notice that Claim $1$ implies that if a longest string which is a subsequence of $X = X_m$, a substring of $Y = Y_n$, and has $P = P_r$ as a subsequence exists then its length is equal to $\max \{\, |Z[i, j, r]| : 1 \leq i \leq m, 1 \leq j \leq n \,\} = \max \{\, M[i][j][r] : 1 \leq i \leq m, 1 \leq j \leq n \,\}$ . 
Hence,  a longest string which is a subsequence of $X$, a substring of $Y$, and has $P$ as a subsequence can be found in the following way. Define one variable called $maxLength$ which eventually represents the length of a 
 longest string which is a subsequence of $X$, a substring of $Y$, and has $P$ as a subsequence and its initial value is $0$. Define another variable called $lastIndexOnY$ which eventually represents the last index of the desired string which is a substring of $Y$ and its initial value
is $n$. Visit all the cells  of $M[i][j][r]$, where $0 \leq i \leq m$ and $0 \leq j \leq n$, in the last two dimensional array created in the algorithm above by using a loop embedded another loop. 
During the visitation, if $M[i][j][r] > maxLength$, then update $maxLength$ and $lastIndexOnY$ as $M[i][j][r]$ and $j$, respectively. After finishing the visitation of all the cells  of $M[i][j][r]$, where $0 \leq i \leq m$ and $0 \leq j \leq n$,  we return
the substring of $Y$ between $(lastIndexOnY - maxLength)$ and $lastIndexOnY$.  \\

The correctness of the above algorithm is ensured by Claim $1$, Claim $2$, and Claim $3$. It is clear that both time complexity and space complexity of the above algorithm are $O((m + 1)(n + 1)(r + 1)) = O(m n r)$. \\

We implemented our algorithm in Java and the program can be found at ``https://sciences.usca.edu/math/\~{}mathdept/rli/ \newline CLCSSeqSStr/CLCSubseqSubstr.pdf".



\end{document}